\documentclass[12pt]{article}
\usepackage[pctex32]{graphics}
\begin{document}
\begin{center} {\bf  Effective Potential on Fuzzy Sphere}
                                                  
\vspace{3cm}

                      Wung-Hong Huang\\
                       Department of Physics\\
                       National Cheng Kung University\\
                       Tainan,70101,Taiwan\\

\end{center}
\vspace{1cm}

The effective potential of quantized scalar field on fuzzy sphere is evaluated to the two-loop level.   We see that  one-loop potential behaves like that in the commutative sphere and the Coleman-Weinberg mechanism of the  radiatively symmetry breaking could be also shown in the fuzzy sphere system.    In the two-loop level, we use the heavy-mass approximation and the high-temperature approximation to perform the evaluations.   The results show that both of the planar and nonplanar Feynman diagrams have inclinations to restore the symmetry breaking in the tree level.   However, the contributions from planar diagrams will dominate over those from nonplanar diagrams by a factor $N^2$.   Thus, at heavy-mass limit or high-temperature system the quantum field on the fuzzy sphere will behave like those on the commutative sphere.  We also see that there is a drastic reduction of the degrees of freedom in the nonplanar diagrams when the particle wavelength is smaller than the noncommutativity scale.

\vspace{2cm}
\begin{flushleft}
E-mail:  whhwung@mail.ncku.edu.tw
\end{flushleft}


\newpage
\section  {Introduction}
   
   Noncommutative fuzzy sphere is known to correspond sphere D2-branes in string theory with background linear B-field [1].  In the presence of  constant RR three-form potential the D0-branes are found to expand into a noncommutative fuzzy sphere configurations [2].  It also knows that the field theories on fuzzy sphere appear naturally from D-brane theory and matrix model with some backgrounds [1,3-6]

      In the ordinary matrix model one could not find the fuzzy sphere solutions.   However, adding a Chern-Simons term  to the matrix model will enable us to describe the noncommutative fuzzy sphere as a classical solution.   Comparing the energy in the various classical solutions one can find that the separated D0-branes will expand into a  largest noncommutative fuzzy sphere to achieve minimum energy [2].

      The problems of the one loop renormalization of the scalar field on the fuzzy sphere have been studied in some recent papers [7-10].    The phenomena of the UL/IR mixing [11,12] on the fuzzy sphere are different from those in the noncommutative plane.   Especially, it has found by Chu. et. al. [9] that ,  in the limit of the commutative sphere, the two point function is regular without UV/IR mixing; however quantization does not commute with the commutative limit, and a finite ``noncommutative anomaly'' survives in the commutative limit.   Chu. et. al. [9] use this to provides an explanation of the UV/IR mixing as an infinite variant of the ``noncommutative anomaly''.

   In this paper we will investigate the effective potential of quantized scalar field [13] on the fuzzy sphere to the two-loop level.   In the section 2, we first review the appearance of the fuzzy sphere in the matrix model with a Chern-Simons term.  (Note that  the Chern-Simons term may be coming from the constant RR three-form potential in the D0-branes system [2].)  We then  quantize the scalar field on the fuzzy sphere and set up the general formula and Feynman rule to evaluate the effective potential.    In the section 3, we see that there is only a single (planar)Feynman diagram  in the one-loop potential.  The potential behaves like that in the commutative system.

   As there appears the Winger $3j$-symbol [14] in Feynman rule we shall adopt some approximations to perform the calculations.  In the section 4, we consider the case in which the mass of scalar field is sufficiently large.  In the section 5, we consider the case in the high temperature environment [15], which  would be that in the early universe.  In both cases the two-loop diagrams are evaluated analytically.   We see that both of the planar Feynman diagrams and nonplanar Feynman diagrams have an inclination to restore the symmetry breaking in the tree level.   Also, the contributions from planar diagrams will dominate over those from nonplanar diagrams by a factor $N^2$.   We also find that the nonplanar diagram will behave as that in the zero space system.   This means that there is a drastic reduction of the degrees of freedom in the nonplanar diagrams when the wavelength of the scalar particle is smaller than the noncommutativity scale [16-18].    As a consequence, the quantum fields on a high-temperature noncommutative fuzzy sphere or a heavy quantized field on a noncommutative fuzzy sphere will behave as that on a commutative sphere.   The last section is devoted to a short conclusion.    

     Note that, historically, Fischler {\it el. al.} [16,17] had first showed that, at high temperature for which the thermal wavelength is smaller than the noncommutativity scale, there is no way to distinguish and count the contributions of modes to the free energy.   Thus, there is a drastic reduction of the degrees of freedom in the non-planar contribution to the thermodynamical potential at high temperature.   In [18] we have also confirmed this property by evaluating the two-loop effective potential on a noncommutative plane.   We in there investigate the same problem  while on a noncommutative fuzzy sphere.   This paper is one of our study concerning the quantum field on the noncommutative geometry [18-20].

\section  {Quantized Scalar Field and Effective Potential on Fuzzy Sphere}
\subsection  {Fuzzy Sphere}
    The matrix model with Chern-Simons term [1] is described by the action:

$$ S= T_0 \ Tr \Big( \frac12 \dot{X_i}^2 + \frac14  [X_i,X_j][X_i,X_j] -
\frac{i}3  \lambda _N \ \epsilon_{ijk} X_i[X_j,X_k] \Big). \eqno{(2.1)}$$
\\
where $X_i, i=1,2,3$ are $N \times N$ matrices and $T_0=\sqrt{2\pi}/g_s$ is
the zero-brane tension.    The static equations of motion are 
$$  [X_j, \bigg( [X_i,X_j] - i\lambda _N\  \epsilon_{ijk} X_k \bigg) ]=0 , \eqno{(2.2)}$$
and the associated energy is 

$$ E =  - T_0 Tr \bigg( \frac {1}{4} [X_i,X_j][X_i,X_j]  - \frac{i}{3} \lambda _N  \epsilon_{ijk} X_i[X_j,X_k] \bigg).  \eqno{(2.3)} $$

   This model admits commutating solutions and static fuzzy sphere solutions [1,2].  The commutating solutions represent $N$ D0 branes and satisfy the relations
     $$[X_i,X_j] =0,   \eqno {(2.4)}$$
which  have the energy
     $$E= 0 .\eqno {(2.5)}$$ 
The static fuzzy sphere solutions satisfy the relations
$$[X_i, X_j] = i \lambda _N \epsilon _{ijk} X_k,  \eqno {(2.6)}$$
and are described by the relations
   $$X_i = \lambda _N  J_i,  \eqno{(2.7)}$$
   $$ X_1^2 + X_2^2 +X_3^2  = R^2. \eqno{(2.8)}$$
where $J_1, J_2, J_3$ define, say, the $N$ dimensional irreducible representation of SU(2) and are labeled by the spin $\alpha= N/2$.   The noncommutativity parameter $\lambda_N$ is of dimension length, and can be taken positive. The radius $R$ define in (2.8) is quantized in units of $\lambda_N$ by
$$\frac {R}{\lambda_N} = \sqrt{\frac{N}{2} \left( \frac{N}{2} +1\right)
}  ;~~~ N = 1,2, ... \eqno{(2.9)}$$

Besides the above solution  $X_i$ may be a direct sum of several irreducible
representation of SU(2).   Such a configuration could also solve the
equation of motion, i.e.
$$ X_i = \lambda_N \oplus_{r=1}^s J_i^{(r)}, ~~ \sum_{r=1}^s (2 j_r + 1)= N . \eqno{(2.10)}$$
The energy E of  these static fuzzy sphere solutions are given by 
$$E = - T_0 \lambda^4 \ \frac{1}{6} \sum_{r=1}^{s} J_r (J_r+1)( 2 J_r +1). \eqno{(2.11)}$$
From the above relation it is clear that the ground state is the N-dimensional  fuzzy sphere [2].

\subsection {Quantized Scalar Field on Fuzzy Sphere: Feynman Rule}

A  field on the fuzzy sphere is defined as an algebra  $S_N^2$ generated by Hermitian operators $\vec {\boldmath X}= (X_1, X_2, X_3)$ which are described in the section 2.1.  The integral of a function $F \in S_N^2$ over the fuzzy sphere is given by [7-10]
$$R^2 \int F  = \frac{4 \pi R^2}{N+1} Tr [ F(X)], \eqno{(2.12)}$$
and the inner product can be defined by 
$$(F_1,F_2) = \int F_1^\dag F_2. \eqno{(2.13)}$$

    A scalar $\Phi^4$ theory on the fuzzy sphere is described by the action
$$S = \int \frac {1}{2} \Phi (\Delta + m^2) \Phi +  \frac{g}{4 !} \Phi^4. \eqno{(2.14)}$$
Here $\Phi$ is a Hermitian field, $m^2$ is the dimensionless  mass square, $g$ is a dimensionless coupling and $\Delta = \sum J_i^2$ is the Laplace operator.    To evaluate the effective potential of the above model we can use the path-integration formulation of Jackiw [13].

   First, we assume that there exists a stationary point at which $\Phi $ is a constant field $\Phi_0$.   Thus 
$$ \left.{\delta S \over \delta \Phi} \right | _ {\phi_0} = 0.\eqno{(2.15)}$$

   Next, we expand the Lagrangian about the stationary point and the action becomes
$$S[\Phi] = S[\Phi_0] + {1\over 2}\int  ~ \tilde{\Phi}(X) ~  \tilde{\Phi}(Y) ~ \left .{\delta^2 S \over {\delta \Phi (X) \delta \Phi(Y)}}  \right | _{\Phi_0}+ \int  ~\tilde {\cal L_I} (\tilde \Phi ,\Phi_0 ), \eqno{(2.16)} $$
\
in which $\tilde{\Phi} \equiv \Phi - \Phi_0 $ and $\tilde {\cal L_I} (\tilde \Phi ,\Phi _0 )$ can be found from  the Lagrangian Eq.(2.14).   

   Then, after expanding  $\tilde \Phi$ in terms of the modes, 
$$\tilde\Phi = \sum_{L,l} a^L_{l} Y^L_l, ~~~  L = 0, 1, ..., N; ~~ -L \leq l \leq L, \eqno{(2.17)}$$
in which the Fourier coefficient $a^L_l$ are treated as the dynamical variables, the path integral quantization procedure [7] is defined by integrating over all possible configuration of  $a^L_l$. Therefore the $k$-points Green's functions are computed by the relation 

$$\langle a^{L_1}_{l_1}   \cdots  a^{L_k}_{l_k} \rangle = \frac{\int [D \tilde\Phi] e^{-S} a^{L_1}_{l_1}   \cdots  a^{L_k}_{l_k}  }{\int [D \tilde \Phi] e^{-S}},\eqno{(2.18)}$$
\\
\
   Note that the complete basis of functions on $S_N^2$ is given by the $(N+1)^2$ spherical harmonics, $Y^L_l, (L = 0, 1, ..., N;  -L \leq l \leq L)$ .  They correspond to the usual spherical harmonics, however the angular momentum has an upper bound $N$ here. This is a characteristic feature of fuzzy sphere. 

   The propagator so obtained is 
$$ D^{-1}(\Phi_0)=\langle a^{L}_{l}   a^{L'}_{l'}{}^\dag \rangle = (-1)^l\langle a^{L}_{l}   a^{L'}_{-l'} \rangle =\delta_{L L'} \delta_{l l'} \frac{1}{L(L+1) + \mu^2 }, \eqno{(2.19)}$$
in which $a^{L}_l{}^\dag = (-1)^l a^L_{-l}$ and
$$\mu^2 = m^2 + {1\over 2}g \Phi_0^2. \eqno{(2.20)}$$
The four-legs vertices are given by
\\
$$\scalebox{1}{\includegraphics{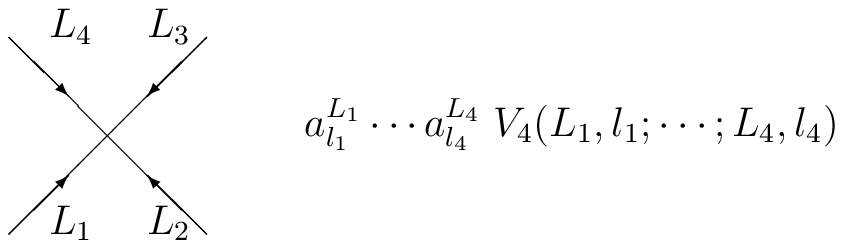}}\eqno{(2.21)}$$
where
$$ V_4(L_1,l_1; \cdots; L_4,l_4) = \frac{g}{4!} \frac{N+1}{4 \pi} 
 (-1)^{L_1+L_2+L_3+L_4} \prod_{i=1}^4 (2L_i+1)^{1/2}  \sum_{L,l}
 (-1)^{l}(2L+1)  \\ $$
$$\times \left( \begin{array}{ccc} L_1&L_2&L\\ l_1&l_2&l \end{array} \right) \left( \begin{array}{ccc} L_3&L_4&L\\ l_3&l_4&-l \end{array} \right) \left\{\begin{array}{ccc} L_1&L_2&L\\ \alpha&\alpha&\alpha \end{array} \right\}  \left\{\begin{array}{ccc} L_3&L_4&L\\ \alpha&\alpha&\alpha \end{array} \right\}. \eqno{(2.22)}$$
\\
The above Feynman rule of propagator and four-legs vertices can be found in [9].   Here we need also the Feynman rule of  three-legs vertices, it is described as
\\
$$\scalebox{1}{\includegraphics{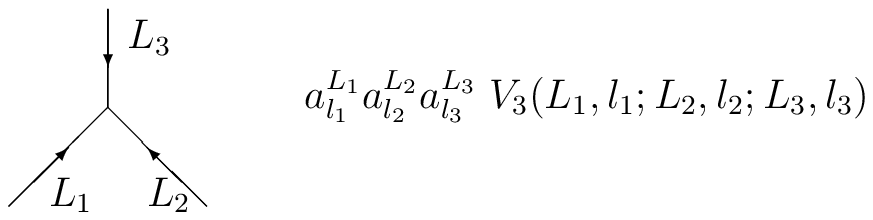}}\eqno{(2.23)}$$
\\
where
$$ V_3(L_1,l_1; L_2,l_2 ; L_3,l_3) = \frac{g}{6} \Phi_0 \sqrt{\frac{N+1}{4\pi}}  (-1)^{2\alpha+L_1+L_2+L_3} \sqrt{(2L_1+1)(2L_2+1)(2L_3+1)} $$
$$~~~~ \times~\left( \begin{array}{ccc} L_1&L_2&L_3\\ l_1&l_2&-l_3 \end{array} \right)  \left\{\begin{array}{ccc} L_1&L_2&L_3\\ \alpha&\alpha&\alpha \end{array} \right\}. \eqno{(2.24)}$$
\\
Here the first bracket is the  Wigner $3j$-symbol and the curly bracket is the $6j$--symbol of $SU(2)$, in the standard mathematical normalization [14].   Note that to derive the above Feynman rule we have used  the following ``fusion'' algebra [14]

$$ Y^I_i Y^J_j =  \sqrt{\frac{N+1}{4\pi}} \sum_{K,k} (-1)^{2\alpha+I+J+K+k} \sqrt{(2I+1)(2J+1)(2K+1)} $$
$$ ~~~~~ \times \left( \begin{array}{ccc} I&J&K\\ i&j&-k \end{array} \right)  \left\{\begin{array}{ccc} I&J&K\\ \alpha&\alpha&\alpha \end{array} \right\} Y^K_k,. \eqno{(2.25)}$$
\\
where the sum is over $0 \leq K \leq N, -K \leq k \leq K$, and $\alpha = N/2$ .

    Then  the effective potential $V(\phi_0)$ is found to be [13]
$$V(\Phi_0) = V_0 (\Phi_0) + {1\over 2} \int \ln det \left[ D^{-1}(\Phi_0)\right] - ~  <exp \left (\int  \tilde {\cal L_I} (\tilde \Phi ,\Phi _0 )\right )>.  \eqno{(2.26)} $$
The first term in Eq.(2.20) is the classical potential.   The second term is
the one-loop contribution from the second term in Eq.(2.16).    The third term  is the higher-loop contribution of the effective potential.   To obtain it we shall evaluate the expectation value of the third term in Eq.(2.16) by the conventional Feynman rule, with $D^{-1}(\Phi_0)$ as the propagator and keep only the connected single-particle irreducible graphs [13].

    In next section, we use the above formula to analyze the one-loop potential, and in the section 4 and 5, we detailed evaluate the two-loop potential on the fuzzy sphere.

\section{One-Loop Effective Potential}

  Now, using the formula in (2.26) the one-loop effective potential  on a fuzzy sphere becomes 

$$  {1\over 2} \int  \ln det \left[ D^{-1}(\Phi_0, {\vec k})\right] =  - {1\over 2}{1\over N+1} \sum_{L,l} \ln \left [L(L+1) + \mu^2\right] $$
$$ = - {1\over 2}{1\over N+1} \sum_{L=0}^N (2L+1) \ln \left[L(L+1) + \mu^2\right]$$
$$ =  - {1\over 2}{1\over N+1} \ln \left ( m^2 + {1\over 2}g \Phi_0^2\right) - {1\over 2} {1\over N+1}\sum_{L=1}^N \ln \left [L(L+1) + m^2 + {1\over 2}g \Phi_0^2\right].\eqno{(3.1)} $$
\\
From the above result we can see the following properties.  In the massless case, we expand the second term around the point $\Phi_0^2 =0$ and it is easily to see that, they will  renormalize the mass and coupling constant and give the high-power terms of $\Phi_0^2$.   However, the first term will become negative  for sufficiently large $\Phi_0$.   Thus it could break the symmetry and the  Coleman-Weinberg mechanism is shown.    In the massive case, we can expand eq.(3.2) around the point $\Phi_0^2 =0$ and it is easily to see that, the one-loop result could only renormalize the mass and coupling constant, and give the high-power terms of $\Phi_0^2$, but they could not change the symmetry property in the tree level.

   The above property is independent of $N$ and thus will behave like that in the commutative sphere system.   Note that as $N \rightarrow \infty$ the fuzzy sphere will become a commutative sphere. 

\section{Two-Loop Effective Potential: Heave Mass Approximation}
    In the two-loop level  we have to evaluate  the following two planar and two nonplanar diagrams.   The contributions of the effective potential from the planar diagram are 
\\
$$\scalebox{1}{\includegraphics{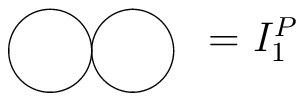}}\hspace{10cm}$$
$$ = ~ - {g\over 4\pi}{1\over3} {1\over (N+1)^2}~\sum_{L} {2L+1\over {L(L+1) + \mu^2}} ~  \sum_{J} {2J+1\over {J(J+1) + \mu^2}} ~~~~~~~~. \eqno{(4.1)} $$
\\
$$\scalebox{1}{\includegraphics{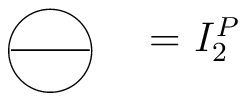}}\hspace{12cm}$$
$$=- \Phi_0^2 ~ ({g\over 4\pi})^2 ~ {1\over 6}~{1\over (N+1)^2}\sum_{L,J} {2L+1\over {L(L+1) +\mu^2}} ~  {2J+1\over {J(J+1)+ \mu^2}} {2L+2J+1\over {(L+J)(L+J+1) +\mu^2}}. \eqno{(4.2)} $$

   The contributions of th effective potential from the nonplanar diagram are
like those in the planar diagram,  while with an extra factor.   They are

$$I_1^N  = ~ - {g\over 4\pi}{1\over6}~{1\over (N+1)^2}\sum_{L,J=0}^N {2L+1\over {L(L+1) +\mu^2}} ~  {2J+1\over {J(J+1) +\mu^2}} $$
$$\times~(-1)^{L+J+2\alpha} ~ (2\alpha +1)\left\{ \begin{array}{ccc} \alpha&\alpha&L\\ \alpha&\alpha&J  \end{array} \right\}. \eqno{(4.3)}$$

 $$I_2^N = - \Phi_0^2 ~ ({g\over 4\pi})^2 ~ {1\over 6} {1\over (N+1)^2}\sum_{L,J} {2L+1\over {L(L+1) +\mu^2}}  {2J+1\over {J(J+1) +\mu^2}} \hspace{5cm}
$$
$$\times ~ {2L+2J+1\over {(L+J)(L+J+1) +\mu^2}} ~ (-1)^{L+J+2\alpha} ~ (2\alpha +1) \left\{ \begin{array}{ccc} \alpha&\alpha&L\\ \alpha&\alpha&J  \end{array} \right\}.  \eqno{(4.4)}$$
~

  To derive the above relations we have use the following identities of the $3j$ and $6j$ symbols, which can be found in [14]: 
The $3j$ symbols satisfy the orthogonality relation
$$\sum_{j,l} \left( \begin{array}{ccc} J&L&K\\ j&l&k \end{array} \right)
\left( \begin{array}{ccc} J&L&K'\\ -j&-l&-k' \end{array} \right)
= \frac{(-1)^{K-L-J}}{2K+1} \delta_{K, K'} \delta_{k, k'}, \eqno{(4.5)}$$
assuming that $(J,L,K)$ form a triangle. 
The $6j$ symbols satisfy the orthogonality relation
$$\sum_{N} (2N+1) \left\{ \begin{array}{ccc} A&B&N\\ C&D&P \end{array} \right\} \left\{ \begin{array}{ccc} A&B&N\\ C&D&Q  \end{array} \right\}
  = \frac 1{2P+1}\; \delta_{P, Q}, \eqno{(4.6)}$$
and the following sum rule 
$$\sum_{N} (-1)^{N+P+Q} (2N+1)  \left\{ \begin{array}{ccc} A&B&N\\ C&D&P  \end{array} \right\} \left\{ \begin{array}{ccc} A&B&N\\ D&C&Q  \end{array} \right\} = \left\{ \begin{array}{ccc} A&C&Q\\ B&D&P  \end{array} \right\},\eqno{(4.7)}$$
assuming that $(A,D,P)$ and $(B,C,P)$ form a triangle. 

   To proceed the calculation we shall adopt some approximations because it is difficult to perform summation over the index in the $6j$ symbols.

   Note that to evaluate the scalar field propagator, Chu et.al. [9] have considered the case that the external moment $L\ll \alpha$.   In this case they used the relation [14] 
$$\left\{ \begin{array}{ccc} \alpha&\alpha&L\\ \alpha&\alpha&J  \end{array} \right\}\approx {(-1)^{L+J+2\alpha \over 2\alpha}}P_L\left(1-{J^2\over 2\alpha^2}\right), ~~~~~~ L\ll \alpha, \eqno{(4.8)}  $$
to evaluate the one-loop correction to the mass square.    They have found that the noncommutative anomaly, which will survive in the commutative sphere limit, is $-{g\over 12\pi}\sum_{k=1}^L{1\over k}$.    

  As we will perform the two-loop calculation in which the summations in (4.3) and (4.4) are over all the values of L, $0\le L\le N$, the approximation of (4.8) can not be used now.   To proceed, we will perform the evaluations of the two-loop diagrams under the approximation that $m^2 \gg N$.

   It is easy to perform the approximation in the planar diagrams.   The results are  

$$I_1^P =  - {g\over 4\pi}{1\over3} {1\over (N+1)^2}~ \sum_{L,J=0}^N {2L+1\over \mu^2} \left(1 - {L(L+1)\over \mu^2}+ \cdot \cdot\cdot\right)  {2J+1\over \mu^2} \left(1 - {J(J+1)\over \mu^2}+ \cdot \cdot\cdot\right) ~$$
$$= - {g\over 4\pi}{1\over3} {1\over (N+1)^2}~ \left[{(N+1)^4\over\mu^4} -{N(N+2)(N+1)^4\over\mu^6}+O\left({N^8\over\mu^8}\right) \right]. \hspace{2cm}\eqno{(4.9)} $$
\\
$$I_2^P=- \Phi_0^2 ~ ({g\over 4\pi})^2 ~ {1\over 6}{1\over (N+1)^2}~\left[{1\over\mu^6} \left({2\over3}N(N+1)^2(N+5)+4N\right) +O\left({N^8\over\mu^8}\right)\right]. \hspace{1cm} \eqno{(4.10)} $$
\\
   The contributions of non-planar diagrams are calculated as follows.
$$I_1^N  = ~ - {g\over 4\pi}{1\over6}{1\over (N+1)^2}~ \sum_{L,J=0}^N {2L+1\over \mu^2} \left(1 - {L(L+1)\over \mu^2}+ \cdot \cdot\cdot\right)  {2J+1\over \mu^2} \left(1 - {J(J+1)\over \mu^2}+ \cdot \cdot\cdot\right) ~$$
$$\times ~ (-1)^{L+J+2\alpha} ~ (2\alpha +1)\left\{ \begin{array}{ccc} \alpha&\alpha&L\\ \alpha&\alpha&J  \end{array} \right\}$$
$$ = - {g\over 4\pi}{1\over6} {1\over (N+1)^2}~ \left[{(N+1)^2\over \mu^4} +O\left({N^6\over\mu^6}\right)\right]. \hspace{5cm}\eqno{(4.11)}$$
\\
To obtain the above results we have used the formula [14]

$$\sum_{L} (2L+1) ~ (-1)^{L+2\alpha} ~ \left\{ \begin{array}{ccc} \alpha&\alpha&L\\ \alpha&\alpha&J  \end{array} \right\} = \delta_{J,0} (2\alpha +1). \eqno{(4.12)}$$
\\
Now using the above formula and a simple relation $(2L+1)(2J+1)(2L+2J+1)= 4(2L+1)J(J+1) +(2L+1)+2L(2L+1)(2J+1)$ we have another result 

 $$I_2^N = - \Phi_0^2 ~ \left({g\over 4\pi}\right)^2 ~ {1\over 6} {1\over (N+1)^2}\sum_{L,J} {2L+1\over \mu^2} \left(1 - {L(L+1)\over \mu^2}+ \cdot \cdot\cdot\right)  {2J+1\over \mu^2} \left(1 - {J(J+1)\over \mu^2}+ \cdot \cdot\cdot\right) $$
$$\times ~   {2L+2J+1\over \mu^2} \left(1 - {(L+J)(L+J+1)\over \mu^2}+ \cdot \cdot\cdot\right)  ~ (-1)^{L+J+2\alpha} ~ (2\alpha +1) \left\{ \begin{array}{ccc} \alpha&\alpha&L\\ \alpha&\alpha&J  \end{array} \right\}$$
$$=  - \Phi_0^2 ~ ({g\over 4\pi})^2 ~ {1\over 6} {1\over (N+1)^2}~ \left[{(N+1)^2\over \mu^6} + O({N^8\over\mu^8})\right].  \hspace{2cm}\eqno{(4.13)}$$
\\
From the above results we find that

$$I_1^N+I_2^N+I_1^P+ I_2^P = - {g\over 4\pi}{1\over6} ~ \left[{1+2(N+1)^2\over \mu^4}+ O\left({N^6\over\mu^6}\right)\right] $$
$$=  {g\over 4\pi}{1\over 6 m^4} ~ \left[2(N+1)^2+1\right] ~ \left(-1+g\Phi_0^2+\cdot\cdot\cdot\right)+ O\left({N^6\over m^6}\right). \eqno{(4.14)}$$
\\
Thus, the two-loop effect potential is to give a positive correction to the mass square term and has an inclination to restore the symmetry breaking in the tree level. (Notice that we set the scale to be $4\pi R^2=1$.)

   Although the heavy mass approximation adopted in this section enables us to evaluate the two-loop effective potential analytically, the approximation is very constricted.   In the next section we will investigate the system in the high temperature environment, which would be that in the early universe and thus is a realistic problem.   Under this approximation we could evaluate the two-loop effective potential analytically and, besides, we find an interesting property of the drastic reduction of the degrees of freedom in the nonplanar diagram.

\section{Two-Loop Effective Potential: High Temperature Approximation}

To consider the finite temperature system the action in (2.14) shall be replaced by 

$$S = \int \frac {1}{2} \Phi (\partial_t^2+ \Delta + m^2) \Phi +  \frac{g}{4 !} \Phi^4. \eqno{(5.1)}$$
\\
The quantized field $\tilde \Phi$ in (2.7) shall be expand in terms of the modes in the following relation 

$$\tilde\Phi = \int_{-\infty}^{\infty} {dp_0 \over (2\pi)} \sum_{L,l}  e^{i {p_0}   t} a^L_l( p_0) Y^L_l, ~~~  L = 0, 1, ..., N; ~~ -L \leq l \leq L, \eqno{(5.2)}$$
\\
in which the Fourier coefficient $a^L_l( p_0)$ are treated as the dynamical variables and  $Y^L_l$ are  the usual spherical harmonics.   The propagator so obtained is 

$$ D^{-1}(\Phi_0)=\langle a(p_0)^{L}_{l}   a(p_0)^{L'}_{l'}{}^\dag \rangle =\delta_{L L'} \delta_{l l'} \frac{1}{p_0^2+L(L+1) + \mu^2 }, \eqno{(5.3)}$$
\\
in which we have let the spacetime to be the Euclidean type.   Then for the theory at finite temperature $T = 1/\beta$ we can  take the following substitutions [15,18]:

 $$p_0 \rightarrow {2 \pi p_0\over \beta },  \eqno{(5.4)}$$
 $$\int d p_0 \rightarrow  {2 \pi \over \beta} \sum_{p_0}. \eqno{(5.5)}$$
\\
in which $p_0$ is an integral.

   Now, we can use the previous prescriptions to evaluate the finite temperature two-loop effective potential.   The contributions of planar diagrams are 

$$I_1^P = - {g\over12\pi} {1\over (N+1)^2}\left({2 \pi \over \beta}\right )^2\sum_{p_0} \left[{(N+1)^4\over[({2 \pi p_0\over \beta})^2+\mu^2]^2} -{N(N+2)(N+1)^4\over[({2 \pi p_0\over \beta})^2+\mu^2]^3} +O\left({N^8\over[({2 \pi p_0\over \beta})^2+\mu^2]^4}\right) \right]$$
$$= - {g\over 12\pi}{(N+1)^2 \over\mu^4} \left({2 \pi \over \beta}\right )^2 + O\left((\beta N)^2\right). \eqno{(5.6)} $$
\\
$$I_2^P=- \Phi_0^2 ~ ({g\over 4\pi})^2 ~ {1\over 6}{1\over (N+1)^2}\left({2 \pi \over \beta}\right)^2\sum_{p_0} \left[{{2\over3}N(N+1)^3(N+5)+4N\over[({2 \pi p_0\over \beta})^2+\mu^2]^3}  +O\left({N^8\over[({2 \pi p_0\over \beta})^2+\mu^2]^4}\right)\right]$$
$$ = - \Phi_0^2 ~ ({g\over 4\pi})^2 ~ {1\over 9}{N(N+1)^2(N+5)+6N\over (N+1)^2~ \mu^6}\left({2 \pi \over \beta}\right)^2 
+  O\left((\beta N)^2\right). \eqno{(5.7)}$$
\\
The contributions of non-planar diagrams are 

$$I_1^N = - {g\over 4\pi}{1\over6} {1\over (N+1)^2}\left({2 \pi \over \beta}\right)^2\sum_{p_0} \left[{(N+1)^2\over [({2 \pi p_0\over \beta})^2+\mu^2]^2} +O\left({N^6\over[({2 \pi p_0\over \beta})^2+\mu^2]^4}\right)\right]$$
$$=- {g\over 24\pi}{1\over\mu^4} \left({2 \pi \over \beta}\right)^2+ O\left((\beta N)^2\right). \eqno{(5.8)}$$
\\
 $$I_2^N = - \Phi_0^2 ~ ({g\over 4\pi})^2 ~ {1\over 6} {1\over (N+1)^2}\left({2 \pi \over \beta}\right)^2\sum_{p_0} \left[{(N+1)^2\over [({2 \pi p_0\over \beta})^2+\mu^2]^3} + O({N^6\over[({2 \pi p_0\over \beta})^2+\mu^2]^4})\right]$$
$$= - \Phi_0^2 ~ ({g\over 4\pi})^2 ~ {1\over 6} {1\over \mu^6}\left({2 \pi \over \beta}\right)^2+O\left((\beta N)^2\right).  \eqno{(5.9)}$$
\\
The above calculations are just those in (4.9), (4.10), (4.11) and (4.13) while added the summation factor (5.5) and replaced $\mu^2$ by ~$({2 \pi p_0\over \beta})^2+\mu^2$.   The terms in $O\left((\beta N)^2\right)$, which may be evaluated by the zeta-function regularization method, become negligible at a sufficiently  high temperature such that $\beta \ll N$.  The results are similar to those in the section 4 and thus both of the planar and nonplanar Feynman diagrams have inclinations to restore the symmetry breaking in the tree level.  

    Comparing the equations (5.6)-(5.9) we see that when $N\gg 1$ the contributions from planar diagrams will dominate over those from nonplanar diagrams by a factor $N^2$.   Thus our results indicate that, at high temperature ($\beta N \ll 1)$ the quantum field on the fuzzy sphere will behave like those on the commutative sphere. (Note that we set the scale to be $4\pi R^2=1$.)

   It is also worthy to notice that the leading contribution from nonplanar diagrams, i.e. (5.8) and (5.9), does not depend on $N$.   This property can be interpreted is as a drastic reduction of the degrees of freedom in the nonplanar diagrams when the thermal wavelength is smaller than the noncommutativity scale.   In fact, the evaluations about the high-temperature two-loop potential at zero space would enable us to see that the potentials are similar to (5.8) and (5.9).  (The calculation is a easy work as there is only a planar diagram.) 

   We now explain the property of reduction of degrees of freedom more clear as following [16-18].   It is known that as $N \rightarrow \infty$ the fuzzy sphere will become a commutative sphere.   Also, when 
$$N \rightarrow \infty,~~~~~ R^2=N\theta /2\rightarrow \infty,~~~~~ keeping~~  \theta~~fixed. \eqno{(5.10)}$$
we will have a noncommutative plane theory [9].   Then, using the definition of  the thermal wavelength $\lambda = \hbar \sqrt {2\pi \beta \over m}$, the high temperature approximation is useful under the condition
     $$1>\beta N ={\lambda^2 m \over 2\pi\hbar^2} N.  \eqno{(5.11)}$$
Now, put back the scale $4\pi R^2(=1)$ and reset the scale to be ${m \over 2\pi\hbar^2} =1$, the above relation becomes $4\pi R^2>\lambda ^2 N$. Finally, using the definition of $\theta$ in (5.10) we see that the high temperature condition becomes. 
$$2\pi \theta > \lambda^2. \eqno{(5.12)}$$
This means that there is a drastic reduction of the degrees of freedom in the nonplanar diagrams when the thermal wavelength is smaller than the noncommutativity scale.

\section {Conclusion}

     We have studied  the effective potential of quantized scalar field on fuzzy sphere to the two-loop level.   We see that one-loop potential will behave like that in the commutative sphere  and  the Coleman-Weinberg mechanism of the  radiatively symmetry breaking could be also shown in the fuzzy sphere system. 

    In the two-loop potential, both of the planar and nonplanar Feynman diagrams can appear.   We use the  heavy-mass approximation and high-temperature approximation to perform the calculations analytically.   We see that both of the planar Feynman diagrams and nonplanar Feynman diagrams have an inclination to restore the symmetry breaking in the tree level.    We also see that the contributions from planar diagrams will dominate over those from nonplanar diagrams by a factor $N^2$. 
  
    We also find that the nonplanar diagram will behave as that in the zero space system.   This means that there is a drastic reduction of the degrees of freedom in the nonplanar diagrams when the thermal wavelength is smaller than the noncommutativity scale.   Thus our results indicate that, at high temperature ($\beta N \ll 1)$ the quantum field on the fuzzy sphere will behave like those on the commutative sphere. 

   In this paper we have only considered the noncommutative scalar field theory.   The more realistic model including the gauge fields on a fuzzy sphere [5,6,21]is deserved studied. However, we believe that the property of the reduction of the degrees of freedom in the nonplanar diagram at high temperature will also be shown in other models.   It remains to be investigated.
  
\newpage
\begin{enumerate}
\item  A.~Y.~Alekseev, A.~Recknagel and V.~Schomerus, ``Non-commutative world-volume geometries: Branes on su(2) and fuzzy sphere'' JHEP {\bf 9909} (1999) 023, [hep-th/9908040]; ``Brane dynamics in background fluxes and non-commutative geometry,'' JHEP {\bf 0005} (2000) 010, [hep-th/0003187].
\item  R.~C.~Myers, ``Dielectric-branes,'' JHEP {\bf 9912} (1999) 022, [hep-th/9910053]. 
\item   Pei-Ming Ho, ``Fuzzy Sphere from Matrix Model'' JHEP {\bf 0012} (2000) 015, [hep-th/0010165]. 
\item  C.~Bachas, M.~Douglas and C.~Schweigert, ``Flux stabilization of D-branes,'' JHEP {\bf 0005} (2000) 048, [hep-th/0003037].
\item K. Hashimoto and K. Krasnov, ``D-brane Solutions in Non-Commutative Gauge Theory on Fuzzy Sphere'', Phys. Rev. {\bf D64}(2001) 046007, [hep-th/0101145]; S. Iso, Y. Kimura, K. Tanaka, and K. Wakatsuki, ``Noncommutative Gauge Theory on Fuzzy Sphere from Matrix Model'', Nucl. Phys. {\bf 604}(2001) 121, [hep-th/0101102].
\item  Y. Kimura, ``Noncommutative Gauge Theory on Fuzzy Sphere and fuzzy torus from Matrix Model'', Prog. theor. Phys. {\bf 106}(2001) 445, [hep-th/0103192].
\item  H.~Grosse, C.~Klimcik and P.~Presnajder, ``Towards finite quantum field theory in noncommutative geometry,'' Int.\ J.\ Theor.\ Phys.\  {\bf 35}, 231 (1996),  [hep-th/9505175].
\item S.~Vaidya, ``Perturbative dynamics on fuzzy $S^2$ and $RP^2$,'' Phys.Lett. B512, 403 (2001), [hep-th/0102212]
\item C. S~Chu, J. Madore and H. Steinacker, ``Scaling Limits of the fuzzy sphere at one loop'', JHEP {\bf 0108}, 038 (2001), [hep-th/0106205].
\item B. P. Dolan, D. O'Connor and P. Presnajder, ``Matrix $\phi ^4$ models on the fuzzy sphere and their continuum limits'',  [hep-th/0109084].
\item S.~Minwalla, M.~Van Raamsdonk and N.~Seiberg, ``Noncommutative perturbative dynamics,'' JHEP {\bf 0002}, 020 (2000), [hep-th/9912072].
\item M.~R.~Douglas and N.~A.~Nekrasov, ``Noncommutative field theory,'' Rev.Mod.Phys. {\bf 73} (2002) 97, [hep-th/0106048].
\item R. Jackiw, Phys. Rev. {\bf D 9} (1974) 1686.
\item D.~A.~Varshalovich, A.~N.~Moskalev and V.~K.~Khersonsky, ``Quantum Theory Of Angular Momentum: Irreducible Tensors, Spherical Harmonics, Vector Coupling Coefficients, 3nj Symbols,'' , Singapore: World Scientific (1988).
\item L. Dolan and R. Jackiw, Phys. Rev. {\bf D 9} (1974) 3320.
\item W. Fischler, E. Gorbatov, A. Kashani-Poor, S. Paban, P. Pouliot and J. Gomis, Evidence for Winding States in Noncommutative Quantum Field Theory, JHEP {\bf 05} (2000) 024, [hep-th/0002067].
\item W. Fischler, E. Gorbatov, A. Kashani-Poor, R. McNees, S. Paban, P. Pouliot and J. Gomis, The interplay between $\theta$ and T, JHEP {\bf 06} (2000) 032, [hep-th/0003216].
\item W. H. Huang, ``High-Temperature Effective Potential of Noncommutative Scalar Field Theory: Reduction of  Degrees of Freedom by Noncommutativity'', Phys. Rev. {\bf D 63} (2001) 125004, [hep-th/0101040]; W. H. Huang, ``Two-loop effective potential in noncommutative scalar field theory'', Phys.Lett. {\bf B496} (2000) 206, [hep-th/0009067].
\item W. H. Huang, ``Finite-Temperature Casimir Effect on the Radius Stabilization of Noncommutative Torus'',JHEP {\bf 0011}, 040 (2000), [hep-th/0011037]; W. H. Huang, ``Casimir Effect on the Radius Stabilization of the Noncommutative Torus'', Phys.Lett. {\bf B497} (2001) 317, [hep-th/0010160].
\item W. H. Huang, ``Quantum Stabilization of Compact Space by Extra Fuzzy Space'', Phys.Lett.{\bf B537} (2002) 311, [hep-th/0203176].
\item U. C.Watamura and S. Watamura, ``Noncommutative Geometry and Gauge Theory on Fuzzy Sphere'', Commun.Math.Phys. {\bf 212} (2000) 395, [hep-th/9801195].

\end{enumerate}
\end{document}